   \documentclass[final,authoryear,5p,times,twocolumn]{elsarticle}
\pdfoutput=1

 \usepackage{graphics}

\usepackage{amssymb}

\journal{Nuclear Instruments and Methods}

\begin{document}

\begin{frontmatter}



\title{Performance of a newly developed SDCCD for X-ray use}


\author[label1]{Hiroshi Tsunemi}
\author[label1]{Shutaro Ueda}
\author[label1]{Kazuo Shigeyama}
\author[label2]{Koji Mori}
\author[label2]{Shoichi Aoyama}
\author[label3]{Shin'ichiro Takagi}
\author{\\ \small{ \underline{(Nucl.Instr.andMeth.A(2010),doi:10.1016/j.nima.2010.08.118)}}}

\address[label1]{Department of Earth and Space Science, Graduate School of Science, Osaka University, 1-1 Machikaneyama, Toyonaka, Osaka 560-0043, Japan}
\address[label2]{University of Miyazaki, 1-1, Gakuen Kibanadai-nishi, Miyazaki 889-2192, Japan}
\address[label3]{Solid State Division, Hamamatsu Photonics K.K., 1126-1, Ichino-cho, Hamamatsu, Shizuoka 435-8558, Japan}

\begin{abstract}
A Scintillator Deposited CCD (SDCCD) is a wide-band X-ray detector consisting of a CCD and a scintillator directly attached to each other.  We assembled the newly developed SDCCD that the scintillator CsI(Tl) is below the fully depleted CCD.  The incident X-rays enter the CCD depletion layer first.  Then, X-rays passing through the depletion layer are absorbed in the CsI(Tl).  The contact surface of the CCD is a back-illuminated side so that we can have good light collection efficiency.  In our experimental setup, we confirmed good performance of our SDCCD detecting many emission lines up to 88\,keV that comes from $^{109}$Cd.  

\end{abstract}


\end{frontmatter}



\section{Introduction}
\label{introduction}

A new type of X-ray detector, SDCCD (Scintillator Deposited CCD) was proposed~\citep{miyata2003}.  It consists of a CCD and a scintillator that are directly attached to each other.  The CCD detects X-rays absorbed in the depletion layer.  Since the CCD is made of silicon wafer, it is difficult to extend the effective energy range above 20\,keV.  The SDCCD employs a fully depleted CCD and is fabricated such that X-rays passing through the depletion layer are absorbed in a scintillator, which extends the CCD energy range up to 100\,keV.

When X-rays are photo-absorbed in the scintillator, they produce some amount of visible photons whose number is proportional to the incident X-ray energy.  Then visible photons are absorbed in the CCD to form an electron cloud.  The conversion gain between the X-ray energy and the number of electrons is quite different between the case that X-rays are photo-absorbed in the depletion layer of the CCD (a CCD event) and the case that X-rays are photo-absorbed in the scintillator (a scintillator event).  The conversion gain of the CCD event is an order of magnitude higher than that of the scintillator event, which depends both on the light yield of the scintillator and on the light collection efficiency of the configuration.  Furthermore, the CCD event forms a very compact electron cloud~\citep{hiraga06} while the scintillator event forms an extended electron cloud.  From this point of view, we have to increase the collection efficiency of visible photons.

The first SDCCD is assembled such that the scintillator was attached to the surface of the front-illuminated (FI) CCD~\citep{miyata2003}.  Since the light collecting efficiency of the FI CCD is limited by the gate structure, the back-illuminated (BI) CCD is suitable to the SDCCD.  From the practical point of view, a BI CCD is employed to fabricate an SDCCD so that they can easily attach the scintillator onto the CCD~\citep{miyata2006}.  The incident X-ray enters into the scintillator first and is converted to visible lights.  In this configuration, all the X-rays are mainly absorbed by the scintillator.  X-rays passing through the scintillator can become CCD events, which is almost no probability to occur.  Therefore, it functions as an X-ray imager whose energy resolution is limited by the scintillator.  This type of the SDCCD is useful to perform a balloon experiment since observable X-rays are limited above 20\,keV.  We performed an experiment, SUMIT~\citep{miyata06spie} employing the combination of the SDCCD and a super mirror that is designed to collect X-rays up to 80\,keV.  The system functioned properly while we failed to recover the system.

Recently, we have fabricated a new type of SDCCD that the X-rays enter into the CCD first then enter into the scintillator so that we can employ this type of the SDCCD in the project of FFAST~\citep{tsunemi08spie}.  We report here the performance of this type of the SDCCD.

\section{Structure and fabrication of the SDCCD}
\label{structure}

\begin{figure}[h]
  \begin{center}
\includegraphics*[width=6cm, bb=0 0 1336 869]{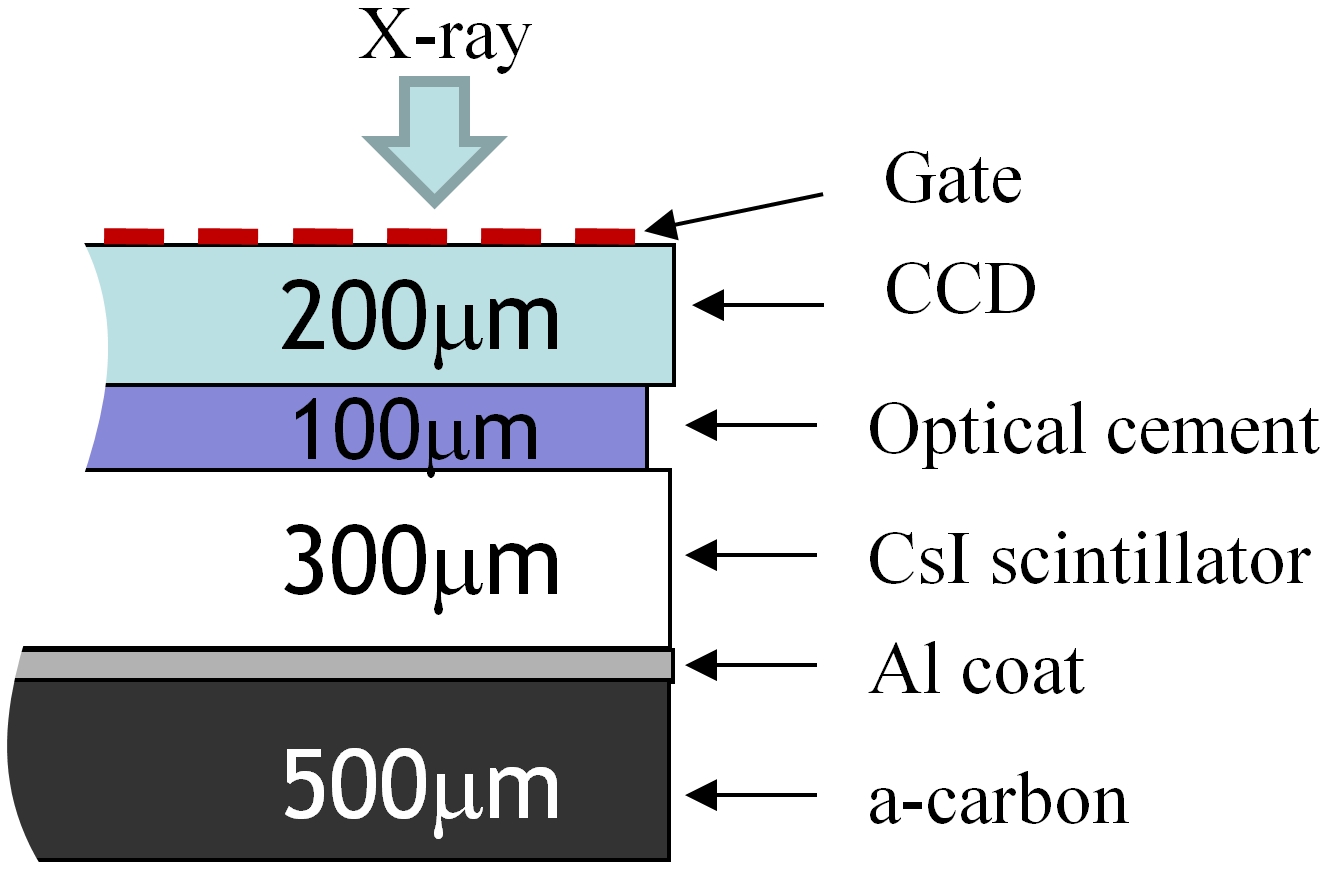}
\end{center}
  \caption{Cross-section of the SDCCD.  We employ a full-depleted back-illuminated CCD.}\label{crosssection}
\end{figure}

Figure \ref{crosssection} shows a schematic cross-section of our SDCCD.  The CCD is placed at the top.  The scintillator is placed below the CCD.  The outlook is identical to that of the FI CCD.  We cannot see the scintillator that is under the CCD.  Since the CCD is working at low temperature, we have to select a scintillator that has high light yield at low temperature~\citep{valentine1993}.  We employed a CsI(Tl) scintillator with a thickness of 300\,$\mu$m that was already employed in the previous SDCCD.  The bottom side of the scintillator has an Al coat so that visible lights are efficiently collected.  This is a columnar shape scintillator with a column size about a few $\mu$m in diameter that reduces the light spread~\citep{tawa2007a}.  We inserted spacers between the CCD wafer and the scintillator so that the gap was kept 100\,$\mu$m.  Then we carefully put optical cement into the gap.

We developed the method of obtaining a good optical contact between the CCD and the CsI(Tl).  Before assembling the SDCCD, we tested the method by using a flat glass instead of the CCD.  The scintillator was about 3cm$\times$6cm and the glass was a little bigger than that.  We visually inspected that our method left no visible bubbles inside the optical cement.  Then, we assembled the SDCCD by using this method.  We have to limit the working temperature of the SDCCD to be --55$^\circ$C that is limited by the temperature allowance of the optical cement.

The CCD employed is called 2K4KCCD that is fabricated in the Hamamatsu photonics K.K.  It is a fully depleted p-channel type CCD and has a depletion layer of 200\,$\mu$m.  The pixel size is 15\,$\mu$m square and the chip size is 3cm$\times$6cm.  When we use it as a frame-transfer mode, the imaging area is 3cm$\times$3cm.  This device is developed with a collaboration of the NAOJ \citep{kamata04, kamata06}.  It is primarily used as a BI CCD that there is no gate structure in the entrance of X-rays (back side).  There is an AR coat for optical use or an Al coat for X-ray use.  In our case, no gate side is attached to the scintillator and the gate side (front side) is the entrance of X-rays.

\begin{figure}[h]
  \begin{center}
\includegraphics*[width=8.5cm, bb=0 0 1407 1027]{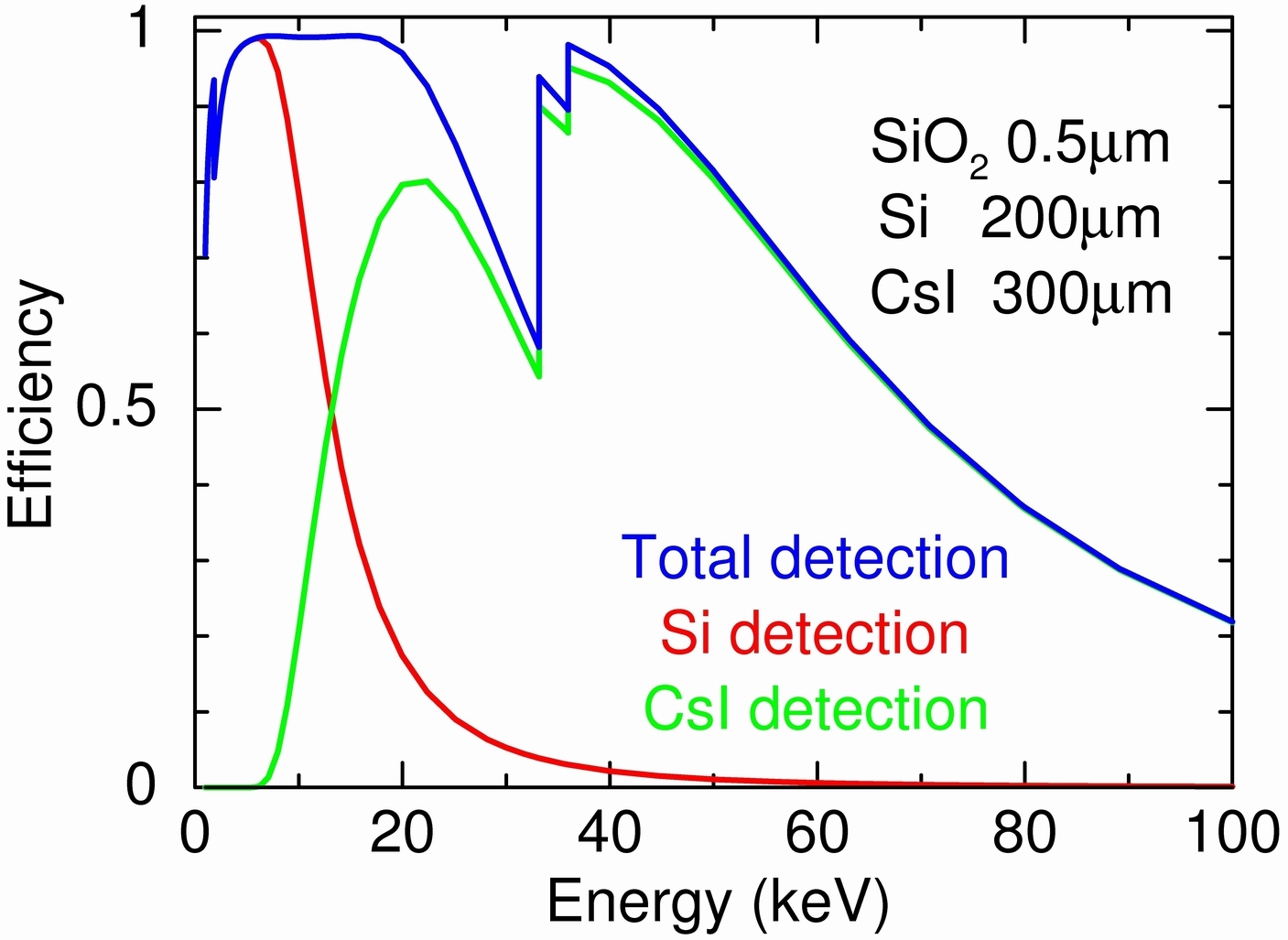}
\end{center}
  \caption{Detection efficiency of the SDCCD}\label{efficiency}
\end{figure}

Figure \ref{efficiency} shows the detection efficiency of our SDCCD.  After passing the gate structure, X-rays below 15\,keV are mainly absorbed in the depletion layer of 200\,$\mu$m Si of which the detection efficiency is shown in red line. X-rays above 15\,keV, passing through the depletion layer and the optical cement, are absorbed in the 300\,$\mu$m CsI(Tl) scintillator of which the detection efficiency is shown in green.  Sum of them is also shown in blue.  Since our CCD is fully depleted, the X-ray absorbed in the CCD forms a primary charge cloud as that in the general X-ray CCD (CCD event).  The charge cloud size is quite small, resulting to be a single pixel event or two-pixel split event.  When it enters into the pixel corner, it may form a four-pixel split.  The X-ray absorbed in the scintillator generates visible lights that enter into the CCD (scintillator event).  Although visible lights form a charge cloud, it is much bigger size than that absorbed in the CCD.  We can expect that the CCD event forms a compact charge size with relatively a large signal while the scintillator event forms a wide-spread charge size with relatively a small signal.  Therefore, we can expect to distinguish them with a signal shape.

\section{Experiment and results}
\label{experiment}

\subsection{Data taking in 16$\times$16 binning mode}
\begin{figure}[h]
  \begin{center}
\includegraphics*[width=9cm, bb=0 0 2283 1404]{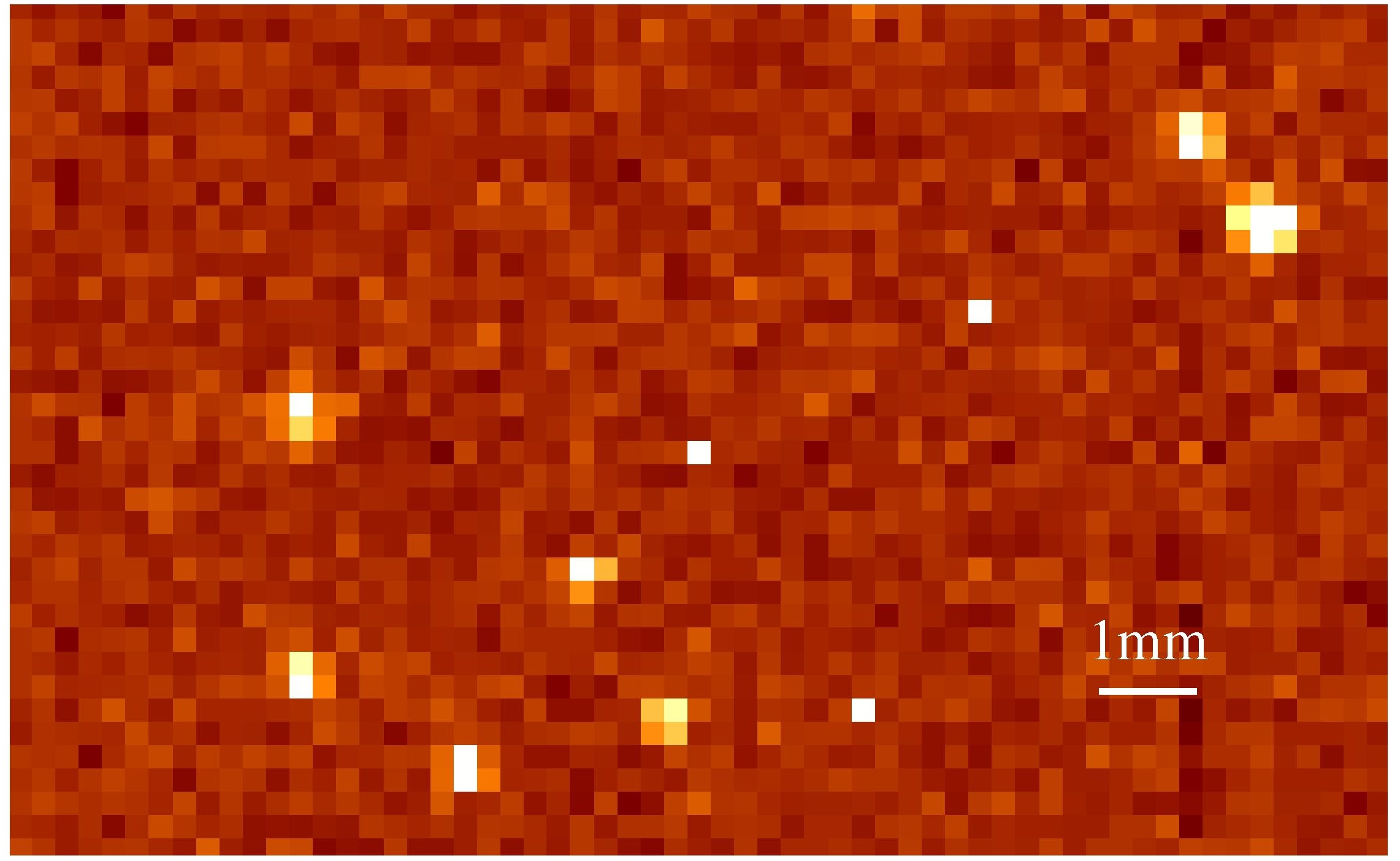}
\end{center}
  \caption{Zoom up the frame image of 16$\times$16 binning.  One pixel corresponds to 240\,$\mu$m square.  Since the density scale is strongly focused near the background level, charge in single pixel events is completely saturated.}\label{16binning_image}
\end{figure}

We mainly employ $^{109}$Cd as a calibration source.  We run the CCD in a various binning mode in which n pixel $\times$ n pixel data are on-chip sum.  In no-binning mode, we can detect CCD events while scintillator events are difficult to be seen.  Then, we increase the number of binning.  In figure \ref{16binning_image}, we show a part of the frame image in 16$\times$16 binning mode.  We can see two types of events.  One is a single pixel event and the other is a split pixel event.  In this mode, the practical pixel size is 240\,$\mu$m square.  Therefore, almost all the CCD events form single pixel events.  Split pixel events (up to 2$\times$2 pixels) by CCD events are expected to be about a few \%.  However, we see many split pixel events, splitting into 3$\times$3 pixels.  

Scintillator events are generated in the CsI(Tl) with a thickness of 300\,$\mu$m.  The spread of the scintillation lights will be an order of its thickness.  Therefore, when we observe them in practical pixel size of 240\,$\mu$m square, they will confine in a 3$\times$3 pixels.  If the spread of signal is within 3$\times$3 pixels, we can easily apply the standard X-ray analysis tools, a grade method, that are developed for the X-ray satellites, ASCA~\citep{tanaka94} and SUZAKU~\citep{mitsuda07}.The grade method is widely employed in the data analysis in X-ray astronomy.  It analyzes the data of 3$\times$3 pixels for each X-ray event and sorts pixels whether or not they exceed the threshold.  There is another method, a fitting method.  The fitting method is to fit the data of 5$\times$5 pixels by a Gaussian profile~\citep{tsuru01} that requires more data and analysis time than those of the grade method.  

\begin{figure}
  \begin{center}
\includegraphics*[width=9cm, bb=10 0 2709 2036]{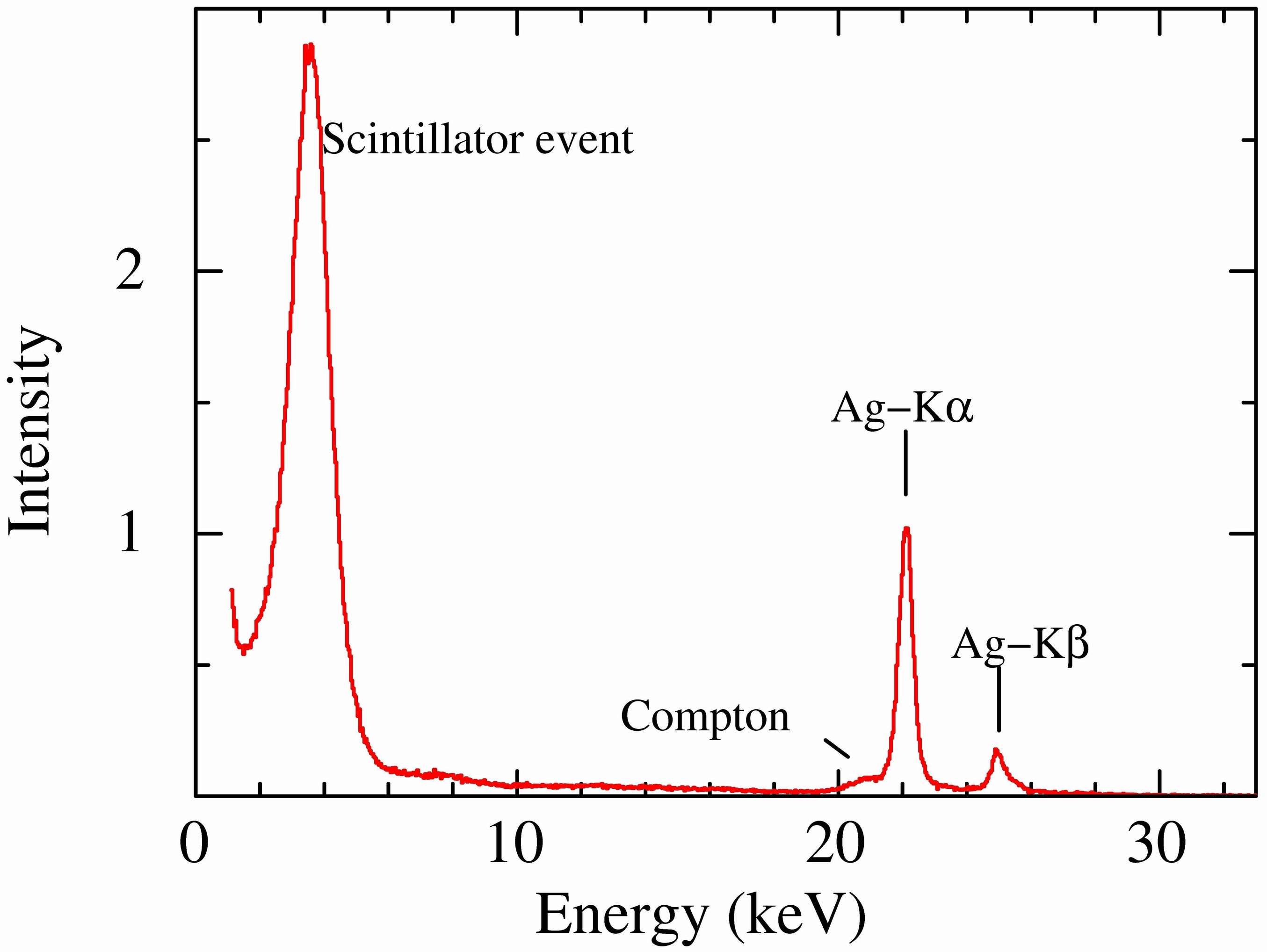}
\end{center}
  \caption{X-ray spectrum from $^{109}$Cd obtained in 16$\times$16 binning mode.  We see X-ray peaks of Ag-K$\alpha$ and Ag-K$\beta$ and a peak around 3.5\,keV that is originated from X-rays absorbed in the CsI(Tl).}\label{16binning_spectrum}
\end{figure}

Figure~\ref{16binning_spectrum} shows a spectrum of X-rays from $^{109}$Cd.  We employed the grade method in the data analysis.  We took events of G0,2,3,4,6 that is the standard criteria in the X-ray spectrum analysis.  Ag-K$\alpha /\beta$ are clearly seen.  Furthermore, we see a peak around 3.5\,keV that is seen only in the SDCCD.  Therefore, the scintillator events yield charge signals that are approximately 1/6th as big as those derived from CCD events.

CCD events can be easily distinguished the Ag-K$\alpha$ from the Ag-K$\beta$.  The apparent intensity of Ag-K$\alpha$ is about 6.7 times larger than that of Ag-K$\beta$.  In the scintillator events, these two X-rays form a single peak due to the poor energy resolution of the CsI(Tl).  We can calculate the number of scintillator events expected.  If we set the thickness of the depletion layer to be 200\,$\mu$m and that of the scintillator to be 300\,$\mu$m, we expect that the number of the scintillator event is 7 times more than that of Ag-K$\alpha$.  Our data show the intensity ratio to be 6.9 that is consistent with our expectation.  

The scintillation light yield of the CsI(Tl) scintillator~\citep{valentine1993} depends on the working temperature, peaking about 60 visible photons/keV at --30$^\circ$C.  It reduces to about 80\% of its peak value at --55$^\circ$C.  The scintillation light from CsI(Tl) shows a peak value around 550\,nm.  The quantum efficiency of the 2K4KCCD is about 85\% at this wave length.  With taking into account these conditions, the expected number of electrons generated inside the CCD is 20--40\, electrons/keV, depending on the reflection efficiency of the Al coat on the back surface of the scintillator.  We obtain the peak of the scintillator events to be 3.5\,keV, which shows that we detected about 40 electrons/keV.  Therefore, we confirm that the Al coat on the bottom of the CsI(Tl) functions properly.

In the 16$\times$16 binning mode, we can employ the grade method that is well established in the X-ray Astronomy.  However, there is a problem in covering a wide energy range.  Our electronics (12-bit ADC) is saturated around 30\,keV/pixel.  If we find a pixel saturated in 3$\times$3 pixels, we will discard the event.  Since the effective pixel size is 240\,$\mu$m, almost all the CCD events generate single-pixel event.  Therefore, the CCD events above 30\,keV cannot be properly analyzed.  Furthermore, the CCD operation of 16$\times$16 binning mode will increase the noise level, particularly at the working temperature of --55$^\circ$C. 

\subsection{Data taking in no binning mode}

\begin{figure}
  \begin{center}
\includegraphics*[width=8cm, bb=0 0 1200 852]{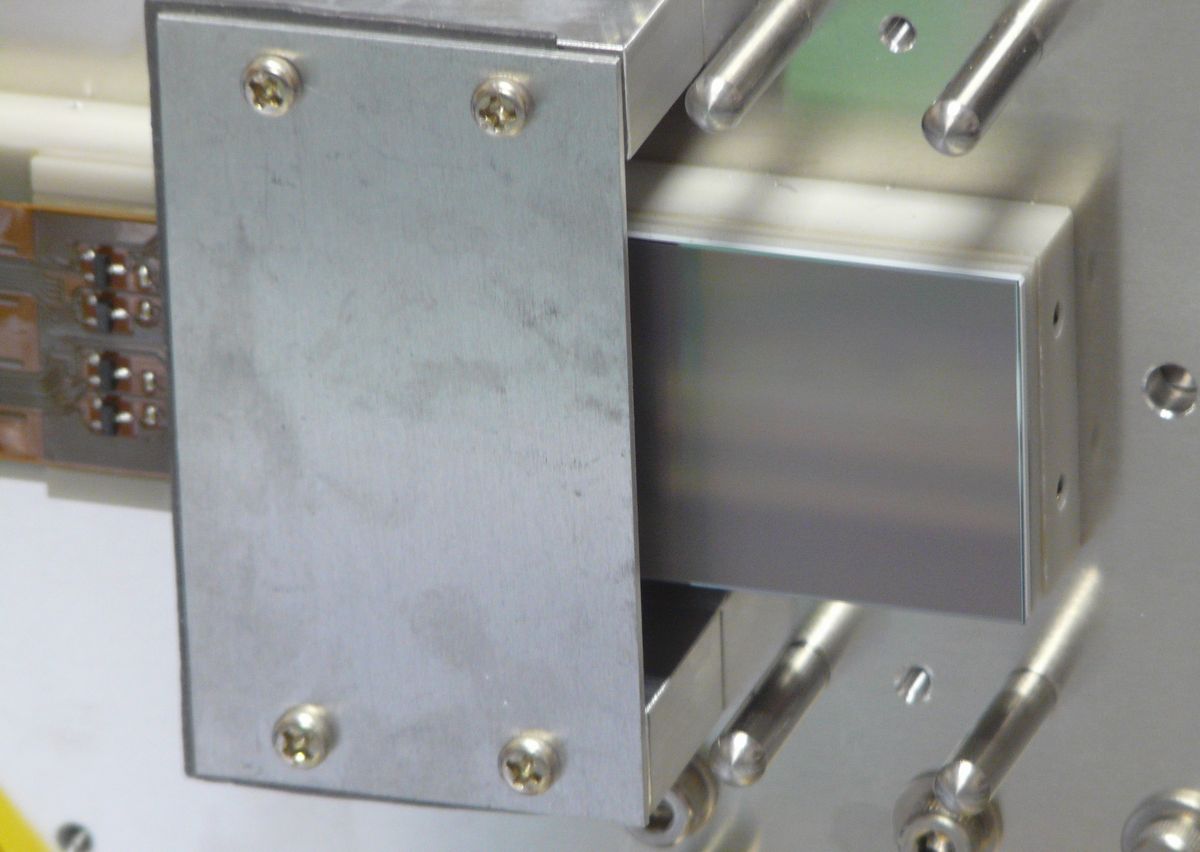}
\end{center}
  \caption{Photo of the SDCCD.  Only the image area is seen.}\label{SDCCDimage}
\end{figure}

SDCCD can detect high energy X-rays while the energy resolution is poor which is limited by the scintillator.  If we can detect X-rays as CCD events, we can achieve a good energy resolution.  Therefore, it is quite useful to detect high energy X-rays as CCD events although the detection efficiency for such a case is low.  In this context, we set up the experimental configuration such that X-rays from $^{109}$Cd source enter into the CCD in a grazing angle.  Furthermore, we restricted the X-ray path by using a metal guide so that we could expect many fluorescence lines.  Figure \ref{SDCCDimage} shows the setup of the SDCCD in which we run it in a frame-transfer mode.

\begin{figure}
  \begin{center}
\includegraphics*[width=9cm, bb=0 0 2813 2029]{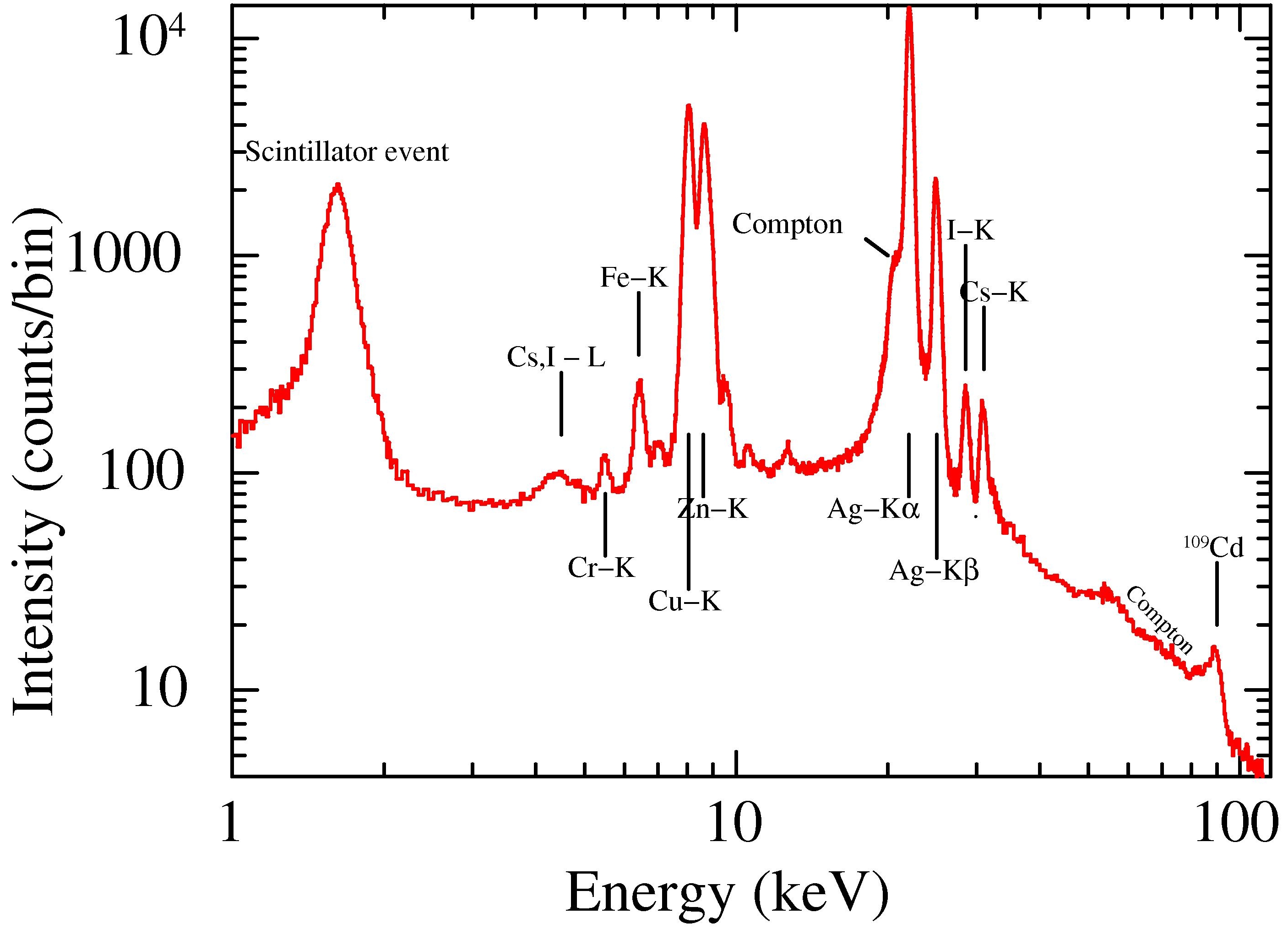}
\end{center}
  \caption{Spectrum obtained by the SDCCD in no-binning mode.  X-rays from $^{109}$Cd generate many secondary X-ray photons.}\label{spectrum}
\end{figure}

We ran the CCD in no binning mode by irradiating X-rays from $^{109}$Cd.  The pixel size is only 15\,$\mu$m square.  We analyzed the data by using a fitting method that can take care of both the compact events and spread events.  Figure~\ref{spectrum} shows the spectrum.  There are many strong emission lines most of which are originated from CCD events.  X-rays of Ag-K$\alpha$ (22.1\,keV) and Ag-K$\beta$ (24.9\,keV) come from the electron capture of $^{109}$Cd.  Emission lines of Cr-K (5.4\,keV), Fe-K (6.4\,keV), Cu-K (8.04\,keV), Zn-K (8.6\,keV) come from the metal guide excited by Ag-K X-rays.  We notice that there is a CCD event at 88\,keV that are nuclear $\gamma$ from $^{109}$Cd.  The detection efficiency of this energy in the depletion layer of the CCD is about 0.1\%.  Most of them penetrate the depletion layer of the CCD and are absorbed in the CsI(Tl).  They excite the CsI and generate I-K (28.5\,keV) and Cs-K (30.8\,keV) X-rays that clearly show the CsI(Tl) scintillator below the CCD.

Since we run the CCD in no binning mode, the high energy X-rays form split pixel events.  The range of the photo-electron in silicon is expressed as $(E/10)^{1.75}\mu$m where $E$ is the electron energy in keV~\citep{janesick85spie}.  Therefore, we can expect the range of photo-electron generated by 88\,keV X-ray to be 3 pixels.  The high energy X-ray will form a split pixel event.  Furthermore, we can expect that each pixel is not saturated.  In this way, we can detect X-rays of 88\,keV in spite that the detection efficiency is very low.  Since the range of the photo-electron generated by 30\,keV X-ray is only half the pixel size, they will be discarded due to the saturation if they form single pixel events.  Only when they form split pixel events, they are properly analyzed and form X-ray peaks.  We should note that the CCD signal has a good linearity up to 88\,keV.

We noticed that there is a peak around 1.7\,keV that is not seen in the conventional CCD.  Therefore, it must come from the scintillator.  Since the scintillator event will extend to 16$\times$16 pixels or more, almost all the pixel signals are well below the detection threshold.  If this peak comes from Ag-K X-rays, the intensity is about 1\% of the expected value in the previous section.  Therefore, almost all the signals are missing due to a large spread of signal.  Furthermore, the peak is only half that is detected in 16$\times$16 binning mode.  Half of the visible lights generated in the CsI(Tl) directly enters into the CCD while the other half transfers to the opposite direction of the CCD and reflects at the Al coat.  The reflected light will widely spread on the CCD and be missing.  Only when the scintillation lights are generated very close to the CCD, direct lights will form relatively compact events that are detected.  This is consistent with that the no-binning data form a peak that is just half that formed in 16$\times$16 binning data.

\subsection{Summary}

We have fabricated a new type of SDCCD.  The incident X-rays enter into a fully depleted FI CCD.  Soft X-rays (below 15\,keV) are photo-absorbed in the depletion layer and form CCD events.  Hard X-rays (above 15\,keV) penetrate the CCD and are photo-absorbed in the CsI(Tl).  Scintillation lights enter into the CCD where they form extended event.

There are two methods of analyzing data: a grade method and a fitting method.  The grade method is well established in the X-ray astronomy.  We have to run the SDCCD in 16$\times$16 binning mode to obtain both the CCD events and the scintillator events where we can apply the grade method.  However, it shows an energy limit for CCD events.  We ran the CCD in no binning mode so that we can apply the fitting method.  This method shows good results for CCD events while it does not work well for scintillator events.  We show that the CCD events has a good linearity up to 88\,keV between the incident X-ray energy and the charge generated in the CCD.  We need to develop an analysis method of taking care of both CCD events and scintillator events simultaneously.

\vspace{5mm}
The development of the SDCCD is supported not only by authors but by many people.  It is partly supported by the Nano-Satellite Research and Development Project in Japan.  K.M. is supported by the ISAS/JAXA program of basic research on satellite-use instruments.

\bibliographystyle{model5-names}
\bibliography{<your-bib-database>}



\end{document}